# Accelerating Pathology Image Data Cross-Comparison on CPU-GPU Hybrid Systems

Kaibo Wang[1]   Yin Huai[1]   Rubao Lee[1]   Fusheng Wang[2,3]   Xiaodong Zhang[1]   Joel H. Saltz[2,3]

[1]Department of Computer Science and Engineering, The Ohio State University
[2]Center for Comprehensive Informatics, Emory University
[3]Department of Biomedical Informatics, Emory University

[1]{wangka,huai,liru,zhang}@cse.ohio-state.edu   [2,3]{fusheng.wang,jhsaltz}@emory.edu

## ABSTRACT

As an important application of spatial databases in pathology imaging analysis, cross-comparing the spatial boundaries of a huge amount of segmented micro-anatomic objects demands extremely data- and compute-intensive operations, requiring high throughput at an affordable cost. However, the performance of spatial database systems has not been satisfactory since their implementations of spatial operations cannot fully utilize the power of modern parallel hardware. In this paper, we provide a customized software solution that exploits GPUs and multi-core CPUs to accelerate spatial cross-comparison in a cost-effective way. Our solution consists of an efficient GPU algorithm and a pipelined system framework with task migration support. Extensive experiments with real-world data sets demonstrate the effectiveness of our solution, which improves the performance of spatial cross-comparison by over 18 times compared with a parallelized spatial database approach.

## 1. INTRODUCTION

Digitized pathology images generated by high resolution scanners enable the microscopic examination of tissue specimens to support clinical diagnosis and biomedical research [10]. With the emerging pathology imaging technology, it is essential to develop and evaluate high quality image analysis algorithms, with iterative efforts on algorithm validation, consolidation, and parameter sensitivity studies. One essential task to support such work is to provide efficient tools for cross-comparing millions of spatial boundaries of segmented micro-anatomic objects. A commonly adopted cross-comparing metric is Jaccard similarity [35], which computes the ratio of the total area of the intersection divided by the total area of the union between two polygon sets.

Building high-performance cross-comparing tools is challenging, due to data explosion in pathology imaging analysis, as in other scientific domains [22, 27]. Whole-slide images made by scanning microscope slides at diagnostic resolution are very large: a typical image may contain over 100,000x100,000 pixels, and millions of objects such as cells or nuclei. A study may involve hundreds of images obtained from a large cohort of subjects. For a large-scale interrelated analysis, there may be dozens of algorithms — with varying parameters — generating many different result sets to be compared and consolidated. Thus, derived data from images of a single study is often in the scale of tens of terabytes, and will be increasingly larger in future clinical environments.

Pathologists mainly rely on spatial database management systems (SDBMS) to execute spatial cross-comparison [36]. However, cross-comparing a huge amount of polygons is time-consuming using SDBMSs, which cannot fully utilize the rich parallel resources of modern hardware. In the era of high-throughput computing, unprecedentedly rich and low-cost parallel computing resources, including GPUs and multi-core CPUs, have been available. In order to use these resources for maximizing execution performance, applications must fully exploit both thread-level and data-level parallelisms and well utilize SIMD (Single Instruction Multiple Data) vector units to parallelize workloads.

However, supporting spatial cross-comparison on a CPU-GPU hybrid platform imposes two major challenges. First, parallelizing spatial operations, such as computing areas of polygon intersection and union, on GPUs requires efficient algorithms. Existing CPU algorithms, e.g., those used in SDBMSs, are branch intensive with irregular data access patterns, which makes them very hard, if not impossible, to parallelize on GPUs. Efficient GPU algorithms, if exist, must successfully exploit massive data parallelisms in the cross-comparing workload and execute them in an SIMD fashion. Second, a GPU-friendly system framework is required to drive the whole spatial cross-comparing workload. The special characteristics of the GPU device require data batching to mitigate communication overhead, and coordinated device sharing to control resource contention. Furthermore, due to the diversity of hardware configurations and workloads, task executions have to be balanced between GPUs and CPUs to maximize resource utilization.

In this paper, we present a customized solution, SCCG (Spatial Cross-comparison on CPUs and GPUs), to address the challenges. Through detailed profiling, we identify that the bottleneck of cross-comparing query execution mainly comes from computing the areas of polygon intersection and union. This explains the low performance of SDBMSs and motivates us to design an efficient GPU algorithm, called PixelBox, to accelerate the spatial operations. Both the design and the implementation of the algorithm are optimized thoroughly to ensure its high performance on GPUs. Moreover, we develop a pipelined system framework for the whole workload, and design a dynamic task migration component to solve the load balancing problem. The pipelined framework has advantages for its natural support of data batching and GPU sharing. The task migration component further improves system throughput by balancing workloads between GPUs and CPUs.

The main contributions of this paper are as follows: 1) PixelBox, an efficient GPU algorithm and its optimized implementation for computing Jaccard similarity of polygon sets; 2) a pipelined framework with task migration support for spatial cross-comparison on a CPU-GPU hybrid platform; and 3) a demonstration of our solution's performance (18x speedup over a parallelized SDBMS) with extensive and intensive experiments using real-world pathology data sets.

The rest of the paper is organized as follows. Section 2 introduces the background and identifies the problem with SDBMSs in processing spatial cross-comparing queries. Our GPU algorithm, PixelBox, is presented in Section 3 to accelerate the bottleneck spatial operations. Section 4 introduces the pipelined framework and the design of a task migration facility for workload balancing. Comprehensive experiments and performance evaluation are presented in Section 5, followed by related works in Section 6 and conclusions in Section 7.

## 2. PROBLEM IDENTIFICATION

### 2.1 Background: Spatial Cross-Comparison

A critical step in pathology imaging analysis is to extract the spatial locations and boundaries of micro-anatomic objects, represented with polygons, from digital slide images using segmentation algorithms [10]. The effectiveness of a segmentation algorithm depends on many factors, such as the quality of microtome staining machines, staining techniques, peculiarities of tissue structures and others. A slight change of algorithm parameters may also lead to dramatic variations in segmentation output. As a result, evaluating the effectiveness and sensitivity of segmentation algorithms has been very important in pathology imaging studies.

The core operation is to cross-compare two sets of polygons, which are segmented by different algorithms or the same algorithm with different parameters, to obtain their degree of similarity. Jaccard similarity, due to its simplicity and meaningful geometric interpretation, has been widely used in pathology to measure the similarity of polygon sets.

Suppose $P$ and $Q$ are two sets of polygons representing the spatial boundaries of objects generated by two methods from the same image. Their Jaccard similarity is defined as

$$J = \frac{\|P \cap Q\|}{\|P \cup Q\|},$$

where $P \cap Q$ and $P \cup Q$ denote the intersection and the union of $P$ and $Q$, and $\|\cdot\|$ is defined as the area of one or multiple polygons in a polygon set. To further simplify the computation, researchers in digital pathology use a variant definition of Jaccard similarity: let $r(p,q) = \frac{\|p \cap q\|}{\|p \cup q\|}$, then

$$J' = \langle \{r(p,q) : p \in P, q \in Q, \|p \cap q\| \neq 0\} \rangle, \quad (1)$$

in which $\langle \cdot \rangle$ represents the average value of all the elements in a set. The greater the value of $J'$ is, the more likely $P$ and $Q$ resemble each other. Compared with $J$, $J'$ does not consider missing polygons that appear in one polygon set but have no intersecting counterpart in the other. Missing polygons can be easily identified by comparing the number of polygons that appear in the intersection with the number of polygons in each polygon set. Other additional measurements of similarity, such as distance of centroids, are omitted in our discussion, as their computational complexity is low.

What makes the computation of $J'$ highly challenging is the huge amount of polygons involved in spatial cross-comparison. Due to the high dependability required by medical analysis, the image base has to be sufficiently large — hundreds of whole slide images are common, with each image generating millions of polygons. Since a single image contains a great number of objects, the average size of polygons extracted from pathology images is usually very small.

To expedite both segmentation and cross-comparing, large image files are usually pre-partitioned into many small tiles so that they can fit into memory and allow parallel segmentations. The generated polygon files for each whole image also reflect the structure of such partitioning: polygons extracted from a single tile are contained in a single polygon file; a group of polygon files constitute the segmentation result for a whole image; different segmentation results for the same image are represented with different groups of polygon files, which are cross-compared with each other for the purpose of algorithm validation or sensitivity studies.

In the rest of the paper, we refer to the area of the intersection of two polygons as *area of intersection*, and the area of the union of two polygons as *area of union*.

### 2.2 Existing Solutions with SDBMSs

Pathologists mainly rely on SDBMSs to support spatial cross-comparison [36]. In this solution, the cross-comparing workflow typically consists of three major steps: first, polygon files (raw data) are loaded into the database; second, indexes are built based on the minimum bounding rectangles (MBRs) of polygons; finally, queries are executed to compute the similarity score. Figure 1(a) shows a cross-comparing query in PostGIS [1] SQL grammar that computes the Jaccard similarity of two polygon sets, named 'oligoastroiii_1_1' and 'oligoastroiii_1_2'. The join condition is expressed with spatial predicate *ST_Intersects*, which tests whether two polygons have intersection. For each pair of intersecting polygons, their area of intersection, area of union, and thus the ratio of the two areas are computed. Spatial operators *ST_Intersection* and *ST_Union* compute the boundaries of the intersection and the union of two polygons, while *ST_Area* returns the area of one or a group of polygons. Finally, these ratios are averaged to derive the similarity score for the whole image.

According to the formula $\|p \cup q\| = \|p\| + \|q\| - \|p \cap q\|$, the query can be re-written so that only the *ST_Intersection* operator is executed for each pair of intersecting polygons, while the area of union can be computed indirectly through the formula. Moreover, *ST_Intersects* can also be removed since we only need records with $ratio > 0$ and whether two

1544

```
SELECT AVG(ratio)
FROM (
  SELECT
    ST_Area(ST_Intersection(p.the_geom, q.the_geom)) /
    ST_Area(ST_Union(p.the_geom, q.the_geom))  AS ratio
  FROM
    oligoastroiii_1_1 AS p, oligoastroiii_1_2 AS q
  WHERE ST_Intersects(p.the_geom, q.the_geom)) AS tmp
WHERE ratio > 0 ;
```

(a) A cross-comparing query without optimizations.

```
SELECT AVG(ratio)
FROM (
  SELECT   area_intersect / (area_p + area_q - area_overlap) AS ratio
  FROM (
    SELECT
      ST_Area(ST_Intersection(p.the_geom, q.the_geom)) AS area_intersect,
      ST_Area(p.the_geom) AS area_p, ST_Area(q.the_geom) AS area_q
    FROM
      oligoastroiii_1_1 AS p, oligoastroiii_1_2 AS q
    WHERE p.the_geom && q.the_geom) AS tmp1
  WHERE area_intersect > 0) AS tmp2 ;
```

(b) A cross-comparing query with optimizations.

**Figure 1:** Cross-comparing queries for the Jaccard similarity of two polygon sets extracted from the same image.

polygons intersect can be determined by their area of intersection. By replacing *ST_Intersects* with the && operator, which tests whether the MBRs of two polygons intersect, we can further optimize the query, as shown in Figure 1(b).

## 2.3 Performance Profiling of SDBMS Solution

To identify the performance bottleneck of cross-comparing queries in SDBMSs, we performed a set of experiments with PostGIS, a popular open-source SDBMS [1] [2]. We used a real-world data set extracted from a brain tumor slide image. The total size of the data set in raw text format is about 750MiB, with two sets of polygons (representing tumor nuclei) each containing over 450,000 polygons, and over 570,000 pairs of polygons with MBR intersections. Details of the platform and the data set will be described in Section 5.

We split the query execution into separate components, and profiled the time spent by the query engine on each component during a single-core execution. The result is presented in Figure 2 for both the unoptimized and optimized queries. Index Search refers to the testing of MBR intersections based on the indexes built. Area_Of_Intersection and Area_Of_Union represent computing the areas of intersection and union, which correspond to the two combo operators, *ST_Area(ST_Intersection())* and *ST_Area(ST_Union())*. ST_Area denotes the other two stand-alone *ST_Area* operators in the optimized query.

For the unoptimized query, *ST_Intersects* (21.8%), *Area_Of_Intersection* (37.4%), and *Area_Of_Union* (36.7%) take the highest percentages of execution time, representing the bottlenecks of the query execution. For the optimized query, since *ST_Intersects* and *Area_Of_Union* are removed from the SQL statement, *Area_Of_Intersection* becomes the sole performance bottleneck, capturing almost 90% of the total query execution time. As the left two bars show, very little time (less than 6%) was spent on index building and index search in both queries. The bar for

---

[1]We also performed similar experiments on a mainstream commercial SDBMS, but its performance was much worse. For simplicity, we only present the results with PostGIS.

[2]Based on our communication with the community of SciDB [2], spatial cross-comparing queries are not natively supported by SciDB.

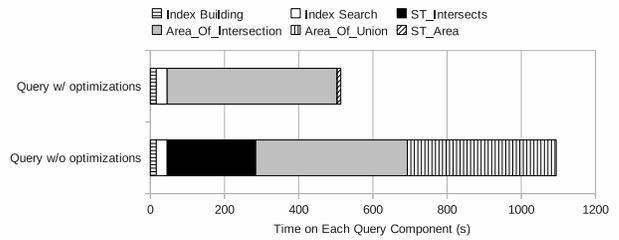

**Figure 2:** Execution time decomposition of cross-comparing queries in PostGIS on a single core.

*ST_Area* shows that the time to compute polygon areas is negligible, and further indicates that the high overhead of *Area_Of_Intersection* and *Area_Of_Union* comes from spatial operators *ST_Intersection* and *ST_Union*.

The profiling result explains the low performance of spatial databases in supporting cross-comparing queries — computing the intersection/union of polygons is too costly as the number of polygon pairs is large. SDBMSs usually rely on some geometric computation libraries, e.g., GEOS [3] in PostGIS, to implement spatial operators. Designed to be *general-purpose*, the algorithms used by these libraries to compute the intersection and union of polygons are compute-intensive and very difficult to parallelize. We analyzed the source codes of respective functions for computing polygon intersection and union in GEOS and another popular geometric library, CGAL [4], and find that only very few sections of codes can be parallelized without significantly changing algorithm structures. Both GEOS and CGAL use generic sweepline algorithms [11], which are not built for computationally intensive queries and thus lead to the limited performance in SDBMSs.

Using a large computing cluster can surely improve system performance. However, unlike in many high-performance computing applications, pathologists can barely afford expensive facilities in real clinical settings [24]. A cost-effective and meanwhile highly productive solution is thus greatly desirable. This motivates us to design a customized solution to accelerate large-scale spatial cross-comparisons. To eliminate the performance bottleneck, our solution needs an efficient GPU algorithm for computing the areas of intersection and union, as will be introduced in the next section.

## 3. THE PIXELBOX ALGORITHM

We describe a GPU algorithm, *PixelBox*, that accepts an array of polygon pairs as input and computes their areas of intersection and union. The design of PixelBox mainly solves three problems: 1) how to parallelize the computation of area of intersection and area of union on GPUs, 2) how to reduce compute intensity when polygon pairs are relatively large, and 3) how to implement the algorithm efficiently on GPUs. We use the terms of NVIDIA CUDA [5] in our description. However, the algorithm design is general and applicable to other GPU architectures and programming models as well.

### 3.1 Pixelization of Polygon Pairs

As measured in the previous section, computing the exact boundaries of polygon intersection/union incurs enormous overhead and has been the main cause to the low performance of SDBMSs in processing cross-comparing queries. However, the most relevant component to the definition of Jaccard similarity (as shown in Formula 1) is the areas, not

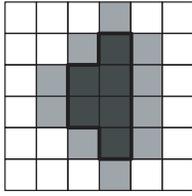

**Figure 3:** Polygons extracted from medical images have axis-aligned edges and integer-valued vertices.

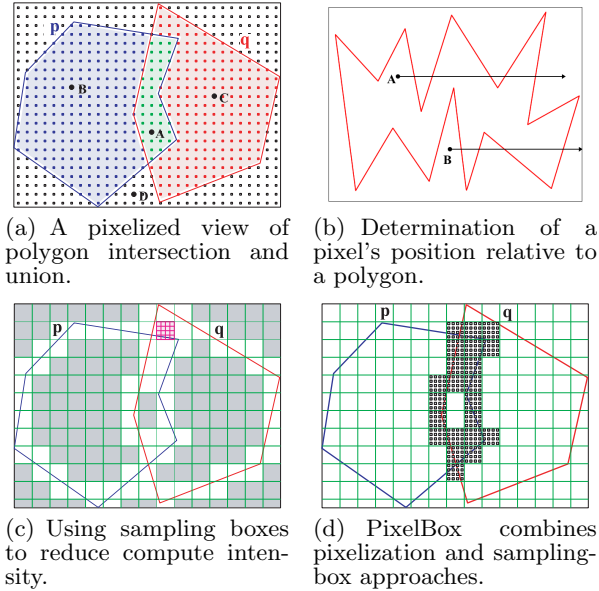

(a) A pixelized view of polygon intersection and union.

(b) Determination of a pixel's position relative to a polygon.

(c) Using sampling boxes to reduce compute intensity.

(d) PixelBox combines pixelization and sampling-box approaches.

**Figure 4:** The principles of PixelBox.

the intermediate boundaries. As a key to enable parallelization on GPUs, PixelBox directly computes the areas without resorting to the exact forms of the intersections or unions.

Polygons extracted from medical images share a common property: *the coordinates of vertices are integer-valued, and the directions of edges are either horizontal or vertical.* This kind of polygons are a special form of rectilinear polygons [34]. As illustrated in Figure 3, since medical images are usually raster images, the boundary of a segmented polygon follows the regular grid lines at the pixel granularity.

Taking advantage of this property, PixelBox treats a polygon as a continuous region surrounded by its spatial boundary on a pixel map. As shown in Figure 4(a), pixels within the MBR of polygons $p$ and $q$ can be classified into three categories: 1) pixels (e.g., $A$) lying inside both $p$ and $q$, 2) pixels (e.g., $B$ and $C$) lying inside one polygon but not the other, and 3) pixels (e.g., $D$) lying outside both. The area of intersection ($\|p \cap q\|$) can be measured by the number of pixels belonging to the first category. The area of union ($\|p \cup q\|$) corresponds to the number of pixels in the first and second categories. Finally, pixels in the third category do not contribute to either $\|p \cap q\|$ or $\|p \cup q\|$. The pixelized view of polygon intersection and union averts the hassle of computing boundaries and, more importantly, exposes a great opportunity for exploiting fine-grained data parallelism hidden in the cross-comparing computation.

In order to determine a pixel's position relative to a polygon, a well-known method is to cast a ray from the pixel and count its number of intersections with the polygon's boundary [28]. As illustrated in Figure 4(b), if the number is odd, the pixel (e.g., $A$) lies inside the polygon; if the number is even, the pixel (e.g., $B$) lies on the outside.

The pixelization method is very suitable for execution on GPUs. Since testing the position of one pixel is totally independent of another, we can parallelize the computation by having multiple threads process the pixels in parallel. Moreover, since the positions of different pixels are computed against the same pair of polygons, the operations performed by different threads follow the SIMD fashion, which is required by GPUs. Finally, the area of intersection and area of union can be computed altogether during a single traversal of all pixels with almost no extra overhead, because the criteria for testing intersection (which uses Boolean AND operation) and union (which uses Boolean OR operation) are both based on each pixel's positions relative to the same polygon pair. As the number of input polygon pairs is large, we can delegate them to multiple thread blocks. For each polygon pair, the contributions of all pixels in the MBR can be computed by all threads within a thread block in parallel.

### 3.2 Reduction of Compute Intensity

The pixelization method described above has a weakness — the compute intensity rises quickly as the number of pixels contained in the MBR increases. Even though polygons are usually very small in pathology imaging applications, as the resolution of scanner lens increases, the sizes of polygons may also increase accordingly to capture more details of the objects. There are also cases when the areas of intersection and union are computed between a small group of relatively large polygons and many small polygons, e.g., when processing an image with a few capillary vessels surrounded by many cells. Moreover, as will be shown in Section 5, even when polygons are small, it is still possible to further bring down the compute intensity and improve performance.

To reduce the intensity of computation and make the algorithm more scalable, PixelBox utilizes another technique, called *sampling boxes*, whose idea is similar to the *adaptive mesh refinement* method [9] in numerical analysis. Due to the continuity of the interior of a polygon, the positions of pixels have *spatial locality* – if one pixel lies inside (or outside) a polygon, other pixels in its neighborhood are likely to lie on the inside (or outside) too, with exceptions near the polygon's boundary. Exploiting this property, we can calculate the areas of intersection and union region by region, instead of pixel by pixel, so that the contribution of all pixels in a region may be computed at once.

This technique is illustrated in Figure 4(c). The MBR of a polygon pair is recursively partitioned into sampling boxes, first at coarser granularity (see the large grid cells in the figure), then going finer at selected sub-regions (e.g., as shown by the small boxes near the top) which need further exploration. For example, when computing the area of intersection, if a sampling box lies completely inside both polygons, the contribution of all pixels within the sampling box is obtained at once, which equals the size of the sampling box; otherwise, the sampling box needs to be partitioned into smaller sub-sampling boxes and tested further. In Figure 4(c), the grey sampling boxes do not need to be further partitioned because their contributions to the areas of intersection and union are already determined.

Similar to the pixelization method, the sampling-box approach requires computing a sampling box's position relative to a polygon, which has three possible values: *inside* – every pixel in the box lies inside the polygon; *outside* – every pixel

1546

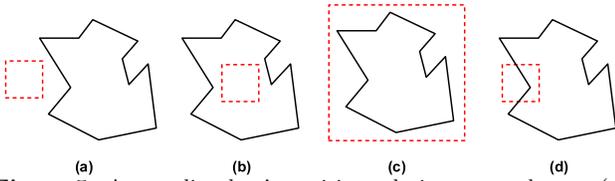

**Figure 5:** A sampling box's position relative to a polygon: (a) outside; (b) inside; (c, d) hover.

in the box lies outside the polygon; and *hover* – some pixels lie inside while others lie outside the polygon.

LEMMA 1. *A sampling box's position relative to a polygon is determined by three conditions: (i) none of the sampling box's four edges crosses through the polygon's boundary; (ii) none of the polygon's vertices lies inside the sampling box; (iii) sampling box's geometric center lies inside the polygon.*

*The sampling box lies inside the polygon if all three conditions are true; it lies outside the polygon if the first two conditions are true but the last is false; it hovers over the polygon in all other cases, when condition (i) or (ii) is false.*

Lemma 1 gives the criteria for computing a sampling box's position, which is further illustrated in Figure 5. For each sampling box, its four edges are tested against the polygon's boundary. If there are edge-to-edge crossings, the sampling box must hover over the polygon (case (d) in Figure 5). Otherwise, if any of the polygon's vertices lies inside the sampling box, the entire polygon must be contained in the sampling box due to the continuity of its boundary, in which case the position is also hover (case (c) in Figure 5); if none of the polygon's vertices is inside the sampling box, the sampling box may be either totally inside (case (b) in Figure 5) or totally outside (case (a) in Figure 5) the polygon, in which case the position of the sampling box's geometric center gives the final answer. If the sampling box's four edges overlap with the polygon's boundary, the sampling box's position can be considered as either inside or outside. The next level of partition will distinguish the contribution of each sub-sampling box to the areas of intersection and union.

Testing the position of a sampling box is more costly than doing this for a pixel. When the granularity of a sampling box is large, the extra overhead is compensated by the amount of per-pixel computations reduced. However, as sampling boxes are more fine-grained, the cost of computing their positions becomes more significant. Moreover, applying sampling boxes requires synchronization between cooperative threads — examination of one sampling box cannot begin until all threads have finished the partitioning of its parent box. Frequent synchronizations lead to low utilization of computing resources and have been one of the main hazards to performance improvement on GPUs [37].

To retain the merits of both efficient data parallelization and low compute intensity, PixelBox combines pixelization with sampling-box techniques. As depicted in Figure 4(d), sampling boxes are applied at first to quickly finish testing for a large number of regions; when the size of a sampling box becomes smaller than a threshold, $T$, the pixelization method takes order and finishes the rest of the computation.

Unlike the pixelization-only method, computing area of intersection and area of union altogether will incur extra overhead with sampling boxes. For example, if a sampling box hovers over one polygon but lies outside the other, its contribution is clear to the area of intersection, but unclear to the area of union; in this case, more fine-grained partitionings are required until the area of union is determined or the pixelization threshold is reached. To reduce the amount of sampling box partitionings and further improve algorithm performance, the area of union is not computed together with the area of intersection in PixelBox. Instead, similar to the query optimization in Figure 1(b), we compute the areas of polygons, and use the formula, $\|p \cup q\| = \|p\| + \|q\| - \|p \cap q\|$, to derive the areas of union indirectly. Computing the area of a simple polygon is very easy to implement on GPUs. With formula[3] $A = \frac{1}{2} \sum_{i=0}^{n-1}(x_i y_{i+1} - x_{i+1} y_i)$, in which $(x_i, y_i)$ is the coordinate of the $i$th vertex of the polygon, we can let different threads compute different vertices and sum up the partial results to get the area.

### 3.3 Optimized Algorithm Implementation

Algorithm 1 shows the pseudocode of PixelBox. Sampling boxes are created and examined recursively — one region is probed from coarser to finer granularities before the next one. A shared stack is used to store the coordinates of the sampling boxes and the flags showing whether each sampling box needs to be further partitioned. For each polygon pair allocated to a thread block, its MBR is pushed onto the stack as the first sampling box (line 13). All threads pop the sampling box on the top of the stack to examine (line 18). If the sampling box does not need to be further probed, all threads will continue to pop the next sampling box (line 19 - 20) until the stack becomes empty and the computation for the polygon pair finishes. For a sampling box that needs to be further examined, if its size is smaller than threshold $T$, the pixelization procedure is applied (line 22 - 28); otherwise, it is partitioned into sub-sampling boxes, and, after further processing, new sampling boxes will be pushed onto the stack by all threads simultaneously (line 30 - 39).

In the algorithm, POLYAREA computes the partial area of a polygon handled by a thread; BOXSIZE returns the number of pixels contained in a sampling box; PIXELINPOLY$(m, i, p)$ computes the position of the $i$th pixel in sampling box $m$ relative to polygon $p$; SUBSAMPBOX$(b, i)$ partitions a sampling box $b$ and returns the $i$th sub-box for a thread to process; BOXPOSITION$(b, p)$ computes the position of sampling box $b$ relative to polygon $p$; BOXCONTINUE computes whether a sampling box needs to be further partitioned based on its positions relative to two polygons; and BOXCONTRIBUTE computes whether a sampling box contributes to the area of intersection according to its position.

The use of a stack to store sampling boxes saves lots of memory space and makes testing sampling box positions and the generation of new sampling boxes parallelized. A synchronization is required before popping a sampling box (line 17) to ensure that thread 0 or the last thread in the thread block has pushed the sampling box to the top of the stack. When threads push new sampling boxes to the stack, they do not overwrite the old stack top (line 37); otherwise, an extra synchronization would be required before pushing new sampling boxes to ensure that the old stack top has been read by all threads. In the current design, the old stack top is marked as 'no further probing' (line 38), and will be omitted by all threads when being popped out again.

The GPU kernel only computes the partial areas of intersections and the partial summed areas of polygons accumulated per thread (lines 5 - 6), which will be reduced later

---
[3]See http://en.wikipedia.org/wiki/Polygon

**Algorithm 1** The PixelBox GPU algorithm.

1: $\{p_i, q_i\}_i$ : the array of input polygon pairs
2: $\{m_i\}_i$ : the MBR of each polygon pair
3: $N$ : total number of polygon pairs
4: $stack[]$ : the shared stack containing sampling boxes
5: $I[N][blockDim.x]$ : partial areas of intersections
6: $A[N][blockDim.x]$ : partial summed areas of polygons
7:
8: **procedure** KERNEL_SAMPBOX
9:     $tid \leftarrow threadIdx.x$
10:     **for** $i = blockIdx.x$ to $N$ **do**
11:         $A[i][tid] \leftarrow A[i][tid] + \text{POLYAREA}(p_i)$
12:         $A[i][tid] \leftarrow A[i][tid] + \text{POLYAREA}(q_i)$
13:         Thread 0: $stack[0] \leftarrow \{m_i, 1\}$
14:         $top \leftarrow 1$
15:         **while** $top > 0$ **do**
16:             $top \leftarrow top - 1$
17:             syncthreads()
18:             $\{box, c\} \leftarrow stack[top]$
19:             **if** $c = 0$ **then**
20:                 **continue**
21:             **else**
22:                 **if** BOXSIZE($box$) $< T$ **then**
23:                     **for** $j \leftarrow tid$ to BOXSIZE($box$) **do**
24:                         $\phi_1 \leftarrow \text{PIXELINPOLY}(box, j, p_i)$
25:                         $\phi_2 \leftarrow \text{PIXELINPOLY}(box, j, q_i)$
26:                         $I[i][tid] \leftarrow I[i][tid] + (\phi_1 \wedge \phi_2)$
27:                         $j \leftarrow j + blockDim.x$
28:                   **end for**
29:                 **else**
30:                     $subbox \leftarrow \text{SUBSAMPBOX}(box, tid)$
31:                     $\phi_1 \leftarrow \text{BOXPOSITION}(box, p_i)$
32:                     $\phi_2 \leftarrow \text{BOXPOSITION}(box, q_i)$
33:                     $c \leftarrow \text{BOXCONTINUE}(\phi_1, \phi_2)$
34:                     $t \leftarrow \text{BOXCONTRIBUTE}(\phi_1, \phi_2)$
35:                     $a \leftarrow (1 - c) \times t \times \text{BOXSIZE}(subbox)$
36:                     $I[i][tid] \leftarrow I[i][tid] + a$
37:                     $stack[top + 1 + tid] \leftarrow \{subbox, c\}$
38:                     Thread 0: $stack[top].c \leftarrow 0$
39:                     $top \leftarrow top + 1 + blockDim.x$
40:                 **end if**
41:             **end if**
42:         **end while**
43:         $i \leftarrow i + gridDim.x$
44:     **end for**
45: **end procedure**

on the CPU to derive the final areas of intersection and union. Reduction is not performed on the GPU because the number of partial values for each polygon pair is relatively small (equal to the thread block size), which makes it not very efficient to execute on the GPU. We measured the time take by the reductions on a CPU core; the cost is negligible compared to other operations on the GPU.

In the rest of this sub-section, we explain some optimizations employed in the algorithm implementation.

**Utilize shared memory**. Effectively using shared memory is important for improving program performance on GPUs [32]. The sampling box stack is frequently read and modified by all threads in a thread block, and thus should be allocated in the shared memory. Meanwhile, polygon vertex data are also repeatedly accessed when computing the positions of pixels and sampling boxes. Loading vertices into shared memory reduces global memory accesses. Due to the limited size of shared memory on GPUs, it is infeasible to allocate for the largest vertex array size. To make a trade off, we set a static size for the shared memory region containing polygon vertices, and only those polygons whose vertices fit into the region are loaded into the shared memory.

**Avoid memory bank conflicts**. Bank conflicts happen when threads in a warp try to access different data items residing in the same shared memory bank simultaneously. In this case, memory access is serialized which decreases both bandwidth and core utilization. In the sampling box procedure, pushing new sampling boxes to the stack may incur bank conflicts if each sampling box is stored continuously in the shared stack. This problem can be solved by separating the stack into five independent ones: four sub-stacks store the coordinates of sampling boxes, and the fifth one stores whether each sampling box needs to be further probed.

**Perform loop unrolling**. Computing pixel or sampling box positions requires comparing with polygon edges in a loop. Unrolling the loop to have multiple polygon edges tested in a single iteration reduces the number of branch instructions and hides memory latency more efficiently.

### 3.4 Related Discussions

**Pixelization threshold** $T$. The pixelization procedure is applied when the number of pixels contained in a sampling box becomes less than the threshold $T$. Let the number of threads in a thread block be $n$, a good value for $T$ should be between $n$ and $n^2$. If $T < n$, the number of pixels contained in the last sampling box is less than the number of threads, which will not keep all threads busy during the pixelization procedure; if $T > n^2$, the last sampling box contains too many pixels, because it could have been further partitioned at least once meanwhile guaranteeing all threads busy during the pixelization procedure. According to our testing (see Section 5.4), $T = \frac{n^2}{2}$ is a very good choice.

**Algorithm accuracy**. Pixelizing polygons may introduce errors into the areas computed. In a general sense, the finer the granularity of pixels is defined, the more accurate the computed result is. For pathology imaging analysis, however, PixelBox does not incur any loss of precision. As explained in Section 3.1, the areas computed equal the numbers of pixels actually lying inside the intersection/union of polygons on the original image. This property generalizes to polygons segmented from any raster image in medical imaging and other applications. We validated the correctness of PixelBox by comparing the areas computed by PixelBox with those computed by PostGIS, and find that the results are the same. We regard the generalization of PixelBox to vectorized polygons as a future work.

**Implications of PixelBox to other spatial operators.** The principal ideas of PixelBox can also be applied to accelerate other compute-intensive spatial operators on GPUs. For example, *ST_Contains* can be implemented by computing the area of intersection and testing whether it equals the area of the object being contained. *ST_Touches* can be accelerated using ideas similar to PixelBox: compare the edges of one polygon with the edges of the other; also test the positions of vertices in one polygon relative to the other polygon; if there is no edge-to-edge crossing, no vertex of one polygon lies within the other polygon, and at least one vertex of one polygon lies on the edge of the other, these two polygons touches each other; otherwise, they do not touch.

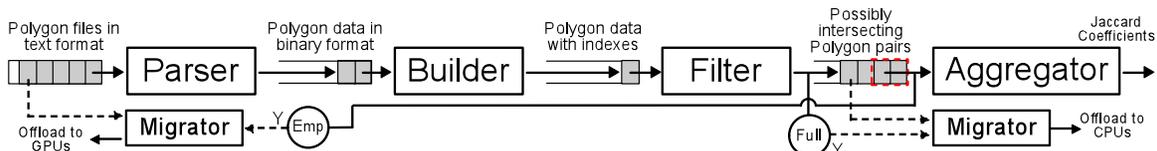
**Figure 6:** A cross-comparing pipeline with dynamic task migrations.

We believe that many frequently used spatial operators in SDBMSs can be parallelized on GPUs by either directly utilizing the PixelBox algorithm or using approaches similar to PixelBox. This is another interesting topic we would like to explore in the future.

## 4. SYSTEM FRAMEWORK

Having presented our core GPU algorithm for computing areas of intersection and union, we are now in a position to introduce how the whole workflow for spatial cross-comparison is implemented and optimized in a CPU-GPU hybrid environment. From the input of the raw text data for polygons to the output of the final results, the workflow consists of multiple logical stages. To fully exploit the rich resources of the underlying CPU/GPU hardware, these stages must be executed in a controllable and dynamically adaptable way. To achieve this goal, the system framework must address three challenges: 1) Since GPU has a disconnected memory space from CPU, input data batching for GPU is needed to compensate the long latency of host-device communication; 2) GPU is an exclusive, non-preemptive compute device [21], thus uncontrolled kernel invocations may cause resource contention and low execution efficiency on GPU; and 3) task executions have to be balanced between CPUs and GPUs in order to maximize system throughput.

In this section, we present our system framework solution. We first introduce our pipelined structure for the whole workload, and then present our dynamic task migration mechanism between CPUs and GPUs.

### 4.1 The Pipelined Structure

We have designed and implemented a pipelined structure for the whole workload. Through inter-stage buffers, task productions and consumptions are overlapped to improve resource utilization and system throughput. As depicted in Figure 6, the cross-comparing pipeline comprises four stages:

1. The *parser* loads polygon files and transforms the format of polygons from text to binaries. This stage executes on CPUs with multiple worker threads.
2. The *builder* builds spatial indexes on the transformed polygon data. Since polygons are small, Hilbert R-Tree [20] is used to accelerate index building. This stage executes on CPUs in a single thread because its execution speed is already very fast.
3. The *filter* performs a pairwise index search on the polygons parsed from every two polygon files, and generates an array of polygon pairs with intersecting MBRs. Similar to the builder, this stage also executes on CPUs with a single worker thread.
4. The *aggregator* computes the areas of intersection and union for each polygon array using our PixelBox algorithm. The ratios of areas are then aggregated to derive the Jaccard similarity for a whole image. Polygon pairs that do not actually intersect, i.e., with the area of intersection being zero, will not be considered.

A computation task at each pipeline stage is defined at the image tile scale. For example, an input task for the parser is to parse two polygon files segmented from the same image tile; an input task for the builder is to build indexes on the two sets of polygons parsed by a single parser task. In practice, a digital image slide may contain hundreds of small image tiles; each tile may contain thousands of polygons. The granularity of tasks defined at image tile level matches the image segmentation procedure, and allows the workload to propagate through the pipeline in a balanced way.

Utilizing such a pipelined framework is critical to solve the aforementioned challenges. First, the work buffers between pipeline stages provide natural support for GPU input data batching. For example, since the number of polygon pairs filtered may be drastically different from tile to tile, it is necessary for the aggregator to group multiple small tasks in its input buffer and send them in a batch to the GPU at once. Second, with a pipelined framework, a single instance of the aggregator consolidates all kernel invocations to the GPUs, which greatly reduces unnecessary contentions and makes the execution more efficient. Finally, the pipelined framework creates a convenient environment for load balancing between CPUs and GPUs, as will be introduced next.

### 4.2 Dynamic Task Migration

Based on the pipelined structure, we have built a task migration component for the whole workflow to achieve load balancing between CPUs and GPUs. First, we have ported the PixelBox algorithms to CPUs (called PixelBox-CPU), and parallelized its execution with multiple worker threads. Second, we have also designed a GPU kernel for the parser stage (called GPU-Parser), whose performance is only comparable to its CPU counterpart since text parsing requires implementing a finite state machine, which has been shown not very efficient for parallel execution [8]. In this way, the parser and the aggregator stages are flexible to execute tasks on both CPUs and GPUs, which creates an opportunity for balancing workload distributions through dynamic task migrations.

What must be noted is that the task migration relies on a special feature of the pipelined framework to detect workload imbalance from the application level. The work buffers between pipeline stages give useful indication on the progress of computation and the status of compute devices. Specifically, if the input buffer of the aggregator stage becomes full, the migrator knows that this stage is making slow progress and the GPUs have been congested. On the other hand, if the input buffer of the aggregator stage becomes empty, it indicates that the GPUs are being under-utilized. In each case, tasks are dynamically migrated from GPUs to CPUs, or from CPUs to GPUs, to mitigate load imbalance and improve system throughput.

To implement the task migration scheme, two background threads, called *migration threads*, are created — one for the aggregator stage, one for the parser stage. They usually stay in the sleeping state and are only woken up when the input buffer of the aggregator stage becomes full or empty. In the case of GPU congestion, the aggregator's migration thread is woken up, which selects the smallest tasks from the input buffer of the aggregator and invokes PixelBox-CPU to execute them. In the case of GPU idleness, the parser's migration thread is woken up to fetch some tasks from the parser's input buffer and execute them on GPUs. The design of the task migration component is also illustrated in Figure 6.

## 5. EXPERIMENTS

This section evaluates our SCCG solution, including the PixelBox algorithm and the system framework. We have implemented PixelBox and GPU-Parser with NVIDIA CUDA 4.0. Intel Threading Building Blocks [25], a popular work-stealing software library for task-based parallelization on CPUs, is used to parallelize text parsing and PixelBox-CPU. The pipelined framework is developed using Pthreads. The dynamic task migration component is built into the execution pipeline, and can be turned on or turned off according to the requirements of respective experiments.

### 5.1 Experiment Methodology

We perform experiments on two platforms. One is a Dell T1500 workstation with an Intel Core i7 860 2.80GHz CPU (4 cores), an NVIDIA GeForce GTX 580 GPU, and 8GiB main memory. The operating system is 64-bit Red Hat Enterprise Linux 6 with 2.6.32 kernel. The other platform is an Amazon EC2 instance with two Intel Xeon X5570 2.93GHz CPUs (totally 8 cores, 16 threads) and two NVIDIA Tesla M2050 GPUs. The size of the main memory is 22 GiB, and the operating system is 64-bit CentOS with 2.6.18 linux kernel. T1500 is primarily used to test the performance of the PixelBox algorithm and the pipelined scheme, and to measure the overall performance of SCCG in cross-comparing all data sets. Amazon EC2 instance is used to measure the performance of a parallelized PostGIS solution to cross-compare the whole data sets. The task migration component is verified on both platforms. The version of PostGIS we used is 1.5.3; the PostgreSQL version is 9.1.3.

Our experiments use 18 real-world data sets extracted from 18 digital pathology images used in a brain tumor research at the authors' institution. The total size of the data sets in raw text format is about 12GiB. The average size of polygons is about 150 in the number of pixels contained, with the standard deviation around 100. The average number of polygons in each data set is about half million, with the largest data set containing over 2 millions.

In all experiments performed in this paper, we do not consider data loading or disk I/O time for the purpose of fair comparison. First, it is well known that the database system has high loading overhead when processing one-pass data with the "first-load-then-query" data processing model. SCCG averts this problem through customized text parsing and pipelined execution to process the polygon stream on the fly. Second, disk I/O, even though still a significant performance factor for SDBMSs, is not longer the severest bottleneck for cross-comparing queries; most time is spent on computation. The effect of disk I/O can be further mitigated through SCCG's pipelined framework by adding a disk pre-fetcher in front of parser stage to sequentially load polygon files into main memory. The use of more advanced storage devices, such as SSDs and disk arrays, can also reduce

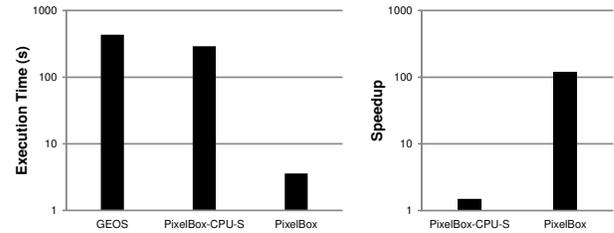
**Figure 7:** Performance comparison of GEOS and PixelBox.

disk I/O time significantly. Thus, in the following experiments, we assume that the polygon data are already loaded into main memory or imported into the database before the pipeline or queries are executed.

### 5.2 Performance of the PixelBox Algorithm

In this subsection, we evaluate the performance of PixelBox and verify some design decisions discussed earlier. The experiments are carried out on the T1500 workstation. Since PostGIS uses GEOS as its geometric computation library, we use the performance of GEOS on a singe core as the baseline in respective experiments. Optimizations similar to the query in Figure 1(b) is used in the baseline to avoid the heavy function call for polygon unions. We select a representative data set, called *oligoastroIII_1*, for the experiments. It contains 462016 polygons in one polygon set and 458878 polygons in the other. Totally, 619609 pairs of polygons, whose MBRs intersect, are filtered.

In Figure 7, we first show the overall performance of GEOS, PixelBox-CPU on a single core (denoted PixelBox-CPU-S), and PixelBox in computing the areas of intersection and union for all 619609 polygon pairs. Both absolute execution times and relative speedups are shown in logarithmic scales. The computation with GEOS takes over 430 seconds. PixelBox-CPU-S performs better than GEOS thanks to algorithm improvement, reducing computation time to about 290 seconds. Compared with GEOS, PixelBox achieves over two-orders-of-magnitude speedup, finishing all computations within only 3.6 seconds. This experiment shows the efficiency of PixelBox algorithm that can fully utilize the power of GPUs to accelerate the computation.

In order to validate several algorithm design decisions, i.e., using sampling boxes to reduce compute intensity, and computing areas of union indirectly, we do a stress testing with PixelBox using a set of 15724 polygon pairs filtered from two representative polygon files in oligoastroIII_1. We increase the polygon sizes by multiplying the coordinates of polygon vertices with a scale factor whose value varies from 1 to 5. The data sets used in this paper are extracted from slide images captured under 20x objective lens. Considering that the resolution of objective lens commonly used is around 40x at the maximum (which increases the sizes of polygons by 4 times), scaling up the coordinates of polygons by a maximum factor of 5 (which increases the sizes of polygons by 25 times) is more than sufficient.

We compare the performance of PixelBox with two base versions: one that uses only the pixelization method (called PixelOnly), the other that combines the pixelization and sampling-box techniques but computes both area of intersection and area of union directly (called PixelBox-NoSep). We tune the grid size, block size, and $T$ (for PixelBox-NoSep and PixelBox), so that all algorithms execute in their best performance. Their execution times are shown in Figure 8.

In all scale factors, the performance of PixelBox-NoSep is

1550

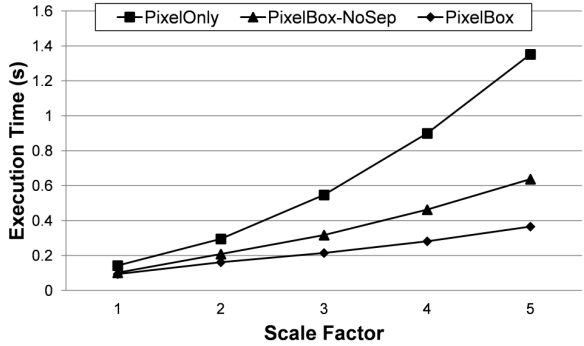

**Figure 8:** Performance of two algorithm decisions: using sampling boxes and computing areas of union indirectly.

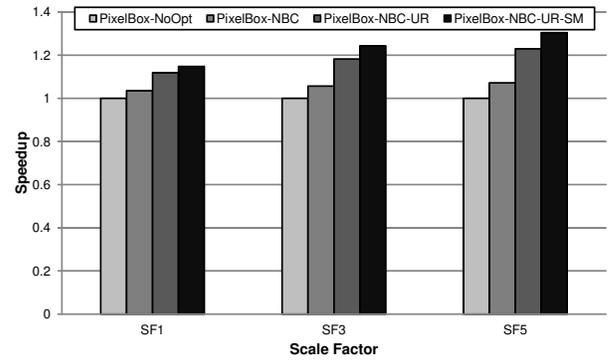

**Figure 9:** Performance impact of various optimization techniques in algorithm implementation.

consistently higher than that of PixelOnly due to the use of sampling boxes, while PixelBox beats the performance of PixelBox-NoSep by further reducing the amount of sampling box partitionings performed. When the scale factor is 1, the overhead of per-pixel examination is relatively low because the sizes of polygons are small. But PixelBox-NoSep and PixelBox still out-perform PixelOnly in this case, reducing execution time by 28% and 34% respectively. As the scale factor increases, the performance of PixelOnly drops rapidly due to the dramatic increase of the number of pixels that must be handled by the algorithm. However, the performance of PixelBox-NoSep and PixelBox only degrades slightly. As the scale factor reaches 5, that is when the sizes of polygons are increased by 25 times, PixelBox-NoSep improves over PixelOnly by reducing execution time by over 50%, while PixelBox shortens the execution time even further by 73% compared with PixelBox-NoSep. This experiment verifies the effectiveness of using sampling boxes to reduce compute intensity. It also shows that, by computing areas of union indirectly, the performance of the algorithm can be further enhanced due to reduced sampling box partitions. It has to be noted that the performance of PixelOnly, PixelBox-NoSep and PixelBox are much higher than the GEOS baseline at all scale factors (it takes GEOS over 11 seconds).

### 5.3 Effectiveness of Optimization Techniques

On the T1500 workstation, we evaluate the effectiveness of various optimization techniques employed during algorithm implementation, i.e., using shared memory (for loading the polygon vertex data), avoiding bank conflicts (when pushing new sampling boxes), and loop unrolling (when computing positions). We take the same set of 15724 polygon pairs used in the previous experiment, with the scaling factors being 1, 3, and 5, and measure the execution times of four variants of the PixelBox algorithm: PixelBox-NoOpt denotes the base version in which none of the optimization techniques are used; PixelBox-NBC denotes the version when bank conflicts are avoided; PixelBox-NBC-UR denotes the version when bank conflicts are avoided and loop unrolling is performed; finally, PixelBox-NBC-UR-SM denotes the version when all optimizations are utilized. In all variants, the sampling box stack is always allocated in shared memory, because otherwise a global heap whose size is proportional to the total number of threads in the whole grid has to be allocated, which we consider an unreasonable design scheme.

The performance of each variant normalized to PixelBox-NoOpt is shown in Figure 9. It can be seen that the optimization techniques discussed above are effective in improving the performance of PixelBox. When the scale factor is 1, the performance is improved by a factor of 1.14 after all optimization techniques are utilized; when the scale factor is 5, the speedup raises to a factor of 1.30. The weights of different optimization techniques to the algorithm performance are, however, varied. The effects of loop unrolling and using shared memory are more significant than that of avoiding bank conflicts. This is because PixelBox spends more time on computing the positions of pixels and sampling boxes than on generating new sampling boxes. Thus, loop unrolling and using shared memory, which improves the efficiency of computing positions, play a larger role in the performance of PixelBox.

### 5.4 Parameter Sensitivity of PixelBox

In order to test the sensitivity of algorithm performance to the pixelization threshold $T$, we take the same set of 15724 polygon pairs used above and measure how the execution time of PixelBox varies as we change the value of $T$. We do the experiments on the T1500 workstation. We set the thread block size to 64, and the performance trend in each scale factor (SF1 to SF5) is shown in Figure 10. The result verifies our analysis for choosing the value of $T$. The performance of PixelBox is sub-optimal when $T$ is too small or too large. It performs the best when the value of $T$ lies between 512 and 4096, which corresponds to the range from $n^2/8$ to $n^2$, in all scale factors. We also repeated the experiment when setting thread block size to other values, and the trend was similar. But when the block size is too large (e.g., $>= 256$), the overall performance of PixelBox degrades. This is because less thread blocks can run concurrently on a multiprocessor and the sampling box parti-tioning will be less fine-grained when block size is too large. According to our experience, setting $n$ to a small value and the value of the pixelization threshold around $n^2/2$ achieves the highest performance.

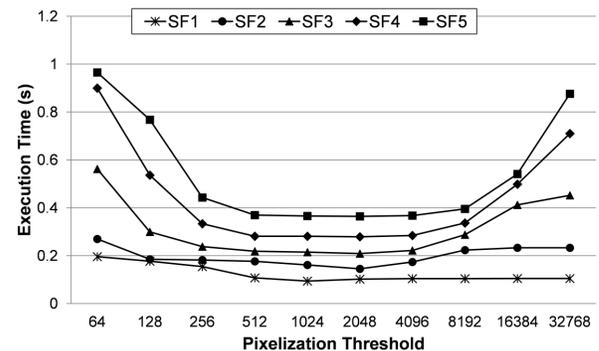

**Figure 10:** The sensitivity of pixelization threshold $T$.

| Scheme | PostGIS-S | NoPipe-S | NoPipe-M | Pipelined |
|--------|-----------|----------|----------|-----------|
| Speedup | 1 | 37.07 | 63.64 | 76.02 |

**Table 1:** Performance comparisons between different schemes.

### 5.5 Performance of the Pipelined Framework

We evaluate the performance of the pipelined framework in this subsection. Task migration is disabled to remove its influence on the pipeline's performance. On the T1500 workstation, we collect the execution times of four schemes that cross-compares the oligoastroIII_1 data set:

- PostGIS-S executes the optimized query shown in Figure 1(b) with PostGIS on a single core;
- NoPipe-S uses a single execution stream that executes a non-pipelined version of the framework in Figure 6, in which the four stages execute sequentially on each pair of input polygon files without pipelining;
- NoPipe-M represents the thread-parallel scheme where multiple execution streams are launched with each one invoking NoPipe-S independently;
- Pipelined is the fully pipelined scheme used in SCCG.

The result is shown in Table 1, with speedup numbers normalized against the PostGIS-S baseline. Since the bottleneck stage of the pipeline has been accelerated by PixelBox on GPUs, NoPipe-S achieves over 37-fold speedup compared with PostGIS-S. NoPipe-M performs better than NoPipe-S (63x speedup over PostGIS-S) because simultaneously issuing multiple streams improves the utilization of resources. However, due to the serialization caused by uncoordinated use of GPUs on the last stage, the CPU resource cannot be well utilized. We measured the CPU utilization during the execution of NoPipe-M and observed that all CPU cores were only about 50% saturated all the times, which confirmed our analysis. The Pipelined scheme achieves the highest performance, accelerating the speed of cross-comparison by a factor of 76 compared with PostGIS-S. The result justifies the use of the pipelined framework and shows the importance of coordination when using GPUs.

### 5.6 Effectiveness of Dynamic Task Migration

In order to verify the design of the task migration component, we perform experiments in three different platform configurations: the T1500 workstation (Config-I), the Amazon EC2 instance with both GPU cards used (Config-II), and the Amazon EC2 instance with only one GPU card used (Config-III). We use the first two configurations to evaluate the effectiveness of the task migration component to offload workloads from CPUs to GPUs, and use the last one for testing load balance in the other direction. Since the GPUs on both platforms are too powerful, in order to make the case of GPU-to-CPU task migrations happen, we purposely slow down PixelBox by selecting a sub-optimal thread block size in Config-III. In real-world system environment, due to concurrent sharing of GPUs with other applications, GPUs may not be exclusively occupied by a single application, which is the case we want to emulate in the last configuration.

The oligoastroIII_1 data set is used in experiments. We show the throughput of task-migration-enabled SCCG normalized to the throughput of task-migration-disabled SCCG

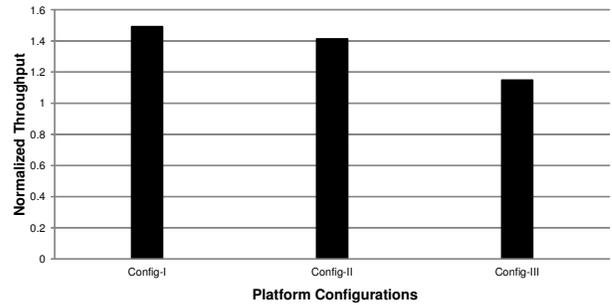

**Figure 11:** Performance benefits of dynamic task migration.

in each configuration. Throughput is defined as the size of data set divided by execution time.

As Figure 11 shows, on T1500 workstation, the throughput of SCCG with dynamic task migration is about 50% higher than SCCG without dynamic task migration. In this setting, the aggregator stage cannot keep the GPU fully occupied, which triggers the migrator to dynamically offload tasks from the parser stage to execute on GPU. This improves the performance of the parser stage and thus enhances the throughput. On Amazon EC2 with both GPUs utilized, the GPU resource still cannot be fully utilized by the aggregator stage. Thus, workloads are migrated from CPUs to GPUs, and the throughput of the pipeline is improved by over 40%. The throughput improvement is lower than Config-I, because the CPUs are more powerful, which causes less workload offloaded to GPUs. On Amazon EC2 with only one GPU utilized, dynamic task migration improves the pipeline throughput by over 14%. In this scenario, the aggregator stage becomes the bottleneck of the pipeline, and some aggregator tasks are migrated to execute on CPUs. But due to the relatively small speed gap between the parser and the aggregator stage and the limited performance of PixelBox-CPU on CPUs, the throughput improvement is smaller compared to other configurations.

### 5.7 Performance Evaluation with All Data Sets

In this section, we give the complete performance results of SCCG compared with a parallelized PostGIS solution over all 18 data sets. The experiments with SCCG are performed on the T1500 workstation with only one GPU card and a 4-core CPU. The experiments with PostGIS are performed on the Amazon EC2 instance with both 4-core CPUs fully utilized. The reason why we choose a less powerful platform for SCCG is to demonstrate both its performance advantage and cost-effectiveness. Query executions in PostGIS are parallelized over all CPU cores by evenly partitioning polygon tables into 16 chunks and launching 16 query streams to process different chunks concurrently. We refer to this execution scheme as PostGIS-M. Being generous to PostGIS, we only consider index building and query execution times; time spent on partitioning polygon tables is not included. We measure the times taken by SCCG and PostGIS-M on cross-comparing each data set, and the relative speedups of SCCG compared with PostGIS-M are presented in Figure 12.

To give an impression on the absolute execution times, it takes PostGIS-M over 1120 seconds to process all data sets, while SCCG finishes all computations within only 64 seconds. As Figure 12 shows, the varied speedups of SCCG over PostGIS-M on different data sets are due to the different numbers and sizes of polygons among the data sets. For example, the first data set contains only 20 polygon files and about 57000 polygons; while the last data set comprises a total of 442 polygon files with over 4 million polygons contained. Among all data sets, SCCG achieves a minimum of 13-fold speedup and a maximum of over 44-fold speedup compared with PostGIS-M. The last column gives the geometric mean of speedups across all data sets, which is over 18 times.

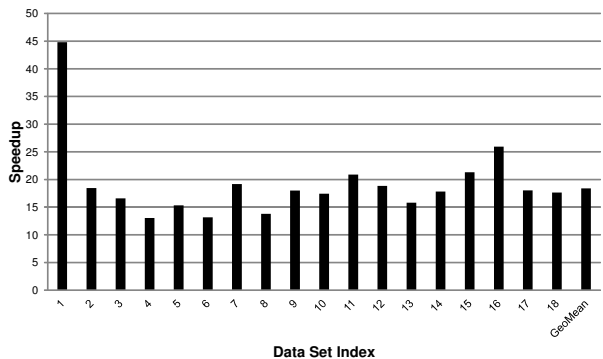

**Figure 12:** The overall performance of SCCG compared with PostGIS-M on 18 data sets.

The result shows the effectiveness of our SCCG solution in improving the performance of spatial cross-comparison at low cost. Two Intel X5570 CPUs cost over $2000, while the total cost of an Intel Core i7 860 CPU and an NVIDIA GTX580 GPU is only about $820 according to the current market price as of March 2012.

## 6. RELATED WORK

Though modern computer architecture has brought rich parallel resources, existing geometric algorithms for spatial operations implemented in the widely used libraries (e.g. CGAL and GEOS) and in major SDBMSs are still single-threaded. There are several attempts of parallel algorithms. A parallel algorithm was proposed in [14] to compute the areas of intersection and union on CPUs. The algorithm was not designed to execute in SIMD fashion, which has been the key to achieve high performance on both CPUs and GPUs in the era of high-throughput computing [26]. As a numerical approximation method, Monte Carlo [13] can be used to compute the areas of intersection and union on GPUs, by repeatedly generating randomized sampling points and counting the number of points lying within the region. However, repeated casting of random sampling points makes Monte Carlo much more compute-intensive than our optimized PixelBox algorithm. A paper [33] proposed to test polygon intersections by drawing polygons on a frame buffer through the OpenGL interfaces and counting the number of pixels with specific colors. This method could be extended to compute the areas of intersection and union, but it would suffer a similar performance problem like the pixelization-only approach due to high compute intensity. The idea of rounding objects to pixels has appeared in fields such as computer graphics [30] and GIS [6], while we realize and utilize the rectilinear property of polygons to solve an important problem in pathology imaging analysis.

Prior work have proposed optimized algorithms and implementations for various database operations on the GPU architecture, including join [18], selection and aggregation [16], sorting [15], tree search [23], list intersection and index compression [7], and transaction execution [19]. Moreover, using a CPU-GPU hybrid environment to accelerate foreign-key joins has been explored in the paper [31]. Compared with these works, we focus on optimizing spatial operations for image comparisons in a CPU-GPU hybrid environment. In addition, considering our system execution framework, related work about the utilization of pipelined execution parallelism can be found in parallel database systems [29] and optimized data-sharing query execution engine [17]. Related work about task scheduling and GPU resource management can be found in work-stealing and real-time systems [12, 21].

## 7. CONCLUSIONS

We have presented our solution for fast cross-comparison of analytical pathology imaging data in a CPU-GPU hybrid environment. After a thorough profiling of a spatial database solution, we identified the performance bottleneck of computing areas of intersection and union on polygon sets. Our PixelBox algorithm and its implementation on GPUs can fundamentally remove the performance bottleneck. Moreover, our pipelined structure with dynamic task migration can efficiently execute the whole workload using CPUs and GPUs. Our solution has been verified through extensive experiments. It achieves more than 18x speedup over parallelized PostGIS when processing real-world pathology data.

We believe our work makes a strong case for performing high-performance, cost-effective digital pathology analysis. The immense power of GPUs and the vectorized functional units on modern hardware must be fully utilized in order to handle the ever-increasing, data-intensive computations. Efficient parallelization of computations on GPUs whilst relies on both the problem characteristics and GPU-optimized algorithm design and implementation. For example, PixelBox trades off a little bit of compute efficiency for a huge gain of data parallelism, and its compute-bound nature also perfectly matches the advantages of GPU architecture. From the system perspective, we consider the incorporation of GPUs into the database ecosystem as an imperative trend with high economic benefits. In a CPU-GPU hybrid environment, many system problems, such as GPU-aware query execution engine, load balancing, and multi-query GPU sharing, need to be addressed.

## 8. ACKNOWLEDGMENTS

This work is supported in part by the National Science Foundation under grants CNS-0834393 and OCI-1147522, CNS 0615155 and CNS 0509326, the National Cancer Institute, National Institutes of Health under contract No. HHSN261200800001E, the National Library of Medicine under grant R01LM009239, and PHS grant UL1RR025008 from the Clinical and Translational Science Award Program, National Institutes of Health, National Center for Research Resources.

## 9. REFERENCES

[1] postgis.refractions.net.
[2] www.scidb.org.
[3] trac.osgeo.org/geos/.
[4] www.cgal.org.
[5] developer.nvidia.com/category/zone/cuda-zone.

[6] resources.esri.com/help//9.3/arcgisengine/java/gp_toolref/conversion_toolbox//converting_features_to_raster_data.htm.

[7] N. Ao, F. Zhang, D. Wu, D. S. Stones, G. Wang, X. Liu, J. Liu, and S. Lin. Efficient parallel lists intersection and index compression algorithms using graphics processing units. *PVLDB*, 4(8):470–481, 2011.

[8] K. Asanovic, R. Bodik, J. Demmel, T. Keaveny, K. Keutzer, J. Kubiatowicz, N. Morgan, D. Patterson, K. Sen, J. Wawrzynek, D. Wessel, and K. Yelick. A view of the parallel computing landscape. *CACM*, 52(10):56–67, 2009.

[9] M. J. Berger and J. Oliger. Adaptive mesh refinement for hyperbolic partial differential equations. *Journal of Computational Physics*, 53(3):484 – 512, 1984.

[10] L. A. D. Cooper, J. Kong, D. A. Gutman, F. Wang, J. Gao, C. Appin, S. Cholleti, T. Pan, A. Sharma, L. Scarpace, T. Mikkelsen, T. Kurc, C. S. Moreno, D. J. Brat, and J. H. Saltz. Integrated morphologic analysis for the identification and characterization of disease subtypes. *Journal of the American Medical Informatics Association*, 19(2):317–323, 2012.

[11] M. de Berg, M. van Kreveld, M. Overmars, and O. Schwarzkopf. *Computational geometry: algorithms and applications*. Springer-Verlag, 1997.

[12] X. Ding, K. Wang, P. B. Gibbons, and X. Zhang. Bws: balanced work stealing for time-sharing multicores. In *EuroSys*, pages 365–378, 2012.

[13] G. S. Fishman. *Monte Carlo: concepts, algorithms, and applications*. Springer-Verlag, 1996.

[14] W. R. Franklin, V. Sivaswami, D. Sun, M. Kankanhalli, and C. Narayanaswami. Calculating the area of overlaid polygons without constructing the overlay. *Cartography and Geographic Information Science*, 21(2):81–89, 1994.

[15] N. Govindaraju, J. Gray, R. Kumar, and D. Manocha. GPUTeraSort: high performance graphics co-processor sorting for large database management. In *SIGMOD*, pages 325–336, 2006.

[16] N. K. Govindaraju, B. Lloyd, W. Wang, M. Lin, and D. Manocha. Fast computation of database operations using graphics processors. In *SIGMOD*, pages 215–226, 2004.

[17] S. Harizopoulos, V. Shkapenyuk, and A. Ailamaki. QPipe: a simultaneously pipelined relational query engine. In *SIGMOD*, pages 383–394, 2005.

[18] B. He, K. Yang, R. Fang, M. Lu, N. Govindaraju, Q. Luo, and P. Sander. Relational joins on graphics processors. In *SIGMOD*, pages 511–524, 2008.

[19] B. He and J. X. Yu. High-throughput transaction executions on graphics processors. *PVLDB*, 4(5):314–325, 2011.

[20] I. Kamel and C. Faloutsos. Hilbert r-tree: an improved r-tree using fractals. In *VLDB*, pages 500–509, 1994.

[21] S. Kato, K. Lakshmanan, R. Rajkumar, and Y. Ishikawa. TimeGraph: GPU scheduling for real-time multi-tasking environments. In *USENIX ATC*, pages 2–2, 2011.

[22] M. L. Kersten, S. Idreos, S. Manegold, and E. Liarou. The researcher's guide to the data deluge: querying a scientific database in just a few seconds. *PVLDB*, 4(12):1474–1477, 2011.

[23] C. Kim, J. Chhugani, N. Satish, E. Sedlar, A. D. Nguyen, T. Kaldewey, V. W. Lee, S. A. Brandt, and P. Dubey. FAST: fast architecture sensitive tree search on modern CPUs and GPUs. In *SIGMOD*, pages 339–350, 2010.

[24] D. B. Kirk and W.-m. W. Hwu. *Programming massively parallel processors: a hands-on approach*. Morgan Kaufmann, 2010.

[25] A. Kukanov and M. J. Voss. The foundations for scalable multi-core software in Intel Threading Building Blocks. *Intel Technology Journal*, 11(4):309–322, 2007.

[26] V. W. Lee, C. Kim, J. Chhugani, M. Deisher, D. Kim, A. D. Nguyen, N. Satish, M. Smelyanskiy, S. Chennupaty, P. Hammarlund, R. Singhal, and P. Dubey. Debunking the 100X GPU vs. CPU myth: an evaluation of throughput computing on CPU and GPU. In *ISCA*, pages 451–460, 2010.

[27] S. Loebman, D. Nunley, Y. Kwon, B. Howe, M. Balazinska, and J. P. Gardner. Analyzing massive astrophysical datasets: can pig/hadoop or a relational dbms help? In *CLUSTER*, pages 1–10, 2009.

[28] J. O'Rourke. *Computational geometry in C*. Cambridge University Press, 1998.

[29] J. Patel, J. Yu, N. Kabra, K. Tufte, B. Nag, J. Burger, N. Hall, K. Ramasamy, R. Lueder, C. Ellmann, J. Kupsch, S. Guo, J. Larson, D. De Witt, and J. Naughton. Building a scaleable geo-spatial DBMS: technology, implementation, and evaluation. In *SIGMOD*, pages 336–347, 1997.

[30] J. Pineda. A parallel algorithm for polygon rasterization. In *SIGGRAPH*, pages 17–20, 1988.

[31] H. Pirk, S. Manegold, and M. L. Kersten. Accelerating foreign-key joins using asymmetric memory channels. In *VLDB - Workshop on Accelerating Data Management Systems Using Modern Processor and Storage Architectures*, pages 585–597, 2011.

[32] S. Ryoo, C. I. Rodrigues, S. S. Baghsorkhi, S. S. Stone, D. B. Kirk, and W.-m. W. Hwu. Optimization principles and application performance evaluation of a multithreaded GPU using CUDA. In *PPoPP*, pages 73–82, 2008.

[33] C. Sun, D. Agrawal, and A. El Abbadi. Hardware acceleration for spatial selections and joins. In *SIGMOD*, pages 455–466, 2003.

[34] L. Surhone, M. Tennoe, and S. Henssonow. *Rectilinear polygon*. Betascript Publishing, 2010.

[35] P.-N. Tan, M. Steinbach, and V. Kumar. *Introduction to Data Mining*. Addison-Wesley Longman Publishing Co., Inc., 2005.

[36] F. Wang, J. Kong, L. Cooper, T. Pan, T. Kurc, W. Chen, A. Sharma, C. Niedermayr, T. Oh, D. Brat, A. Farris, D. Foran, and J. Saltz. A data model and database for high-resolution pathology analytical image informatics. *Journal of Pathology Informatics*, 2(1):32, 2011.

[37] Y. Zhang and J. D. Owens. A quantitative performance analysis model for GPU architectures. In *HPCA*, pages 382–393, 2011.